\documentclass[prl,twocolumn,preprintnumbers,superscriptaddress,showpacs]{revtex4-2}
\usepackage[colorlinks=true, pdfstartview=FitV, linkcolor=red, citecolor=blue, urlcolor=blue, pdftitle={},pdfsubject={}, pdfkeywords={}]{hyperref}

\newcommand\sect[1]{{\it #1.}---}

\usepackage{stmaryrd}
\usepackage{comment}
\usepackage{color}
\usepackage{amsmath}
\usepackage{amssymb}
\usepackage{amsthm}
\usepackage{mathrsfs}
\usepackage{graphicx}
\usepackage{marvosym}
\usepackage{amsmath,bm}
\usepackage{array}
\usepackage{latexsym}
\usepackage{enumerate}
\usepackage{slashed}
\usepackage{hyperref}
\usepackage{color}

\allowdisplaybreaks[1]

\usepackage[normalem]{ulem}

\begin{document}
\title{Anomalous spin polarization from turbulent color fields}
\author{Berndt M\"uller}
\affiliation{
Department of Physics, Duke University, Durham, North Carolina 27708, USA.\\
}
\author{Di-Lun Yang}
\affiliation{
Institute of Physics, Academia Sinica, Taipei, 11529, Taiwan\\
}
\begin{abstract}
We study the important, yet widely overlooked, role of gluons for spin transport with a connection to local parity violation in quark gluon plasmas. We employ the formalism of quantum kinetic theory to quarks in weakly coupled quantum chromodynamics to derive the source terms for quark spin polarization. These source terms involve parity-odd correlators of dynamically generated color fields in near-equilibrium quark gluon plasmas and give rise to locally fluctuating axial charge currents. Our results provide a possible explanation for the spin alignment of vector mesons measured in high-energy nuclear collisions.          
\end{abstract}

\maketitle

\sect{Introduction}
Recent observations of global spin polarization of hadrons in relativistic heavy ion collisions \cite{STAR:2017ckg,STAR:2019erd,ALICE:2019aid,Singha:2020qns} have inspired theoretical studies aimed at understanding the origin of the polarization and the spin transport of partons in quark gluon plasmas (QGP). It is generally believed that the large angular momentum present in peripheral collisions results in spin polarization of quarks and gluons through spin-orbit interaction, and that this polarization is inherited by hadrons after hadronization and freeze-out \cite{Liang:2004ph}. When hadrons or constituent quarks reach thermal equilibrium globally, their polarization is dictated by thermal vorticity \cite{Becattini2013a,Fang:2016vpj}, while there exist further corrections in local equilibrium \cite{Hidaka:2017auj,Liu:2020dxg,Liu:2021uhn,Becattini:2021suc} (see Refs.~\cite{Becattini:2013vja,Fu:2021pok,Becattini:2021iol,Yi:2021ryh,Ryu:2021lnx} for related simulations). Nevertheless, in high-energy collisions, the large angular momentum is no longer effectively deposited without nuclear stopping, and the vorticity drops \cite{Jiang:2016woz,Deng:2016gyh}. The substantial spin alignments of vector mesons observed in LHC \cite{ALICE:2019aid} may require alternative mechanisms for explanations (see e.g. Refs.~\cite{Sheng:2019kmk,Sheng:2020ghv,Xia:2020tyd,Goncalves:2021ziy}).

These phenomenological studies have been supplemented by more formal investigations of quantum kinetic theory (QKT) \cite{Gao:2019znl,Weickgenannt:2019dks,Hattori:2019ahi,Wang:2019moi,Li:2019qkf,Yang:2020hri,Weickgenannt:2020aaf,Wang:2020pej,Hattori:2020gqh} and hydrodynamics for spinning fluids \cite{Montenegro:2017rbu,Florkowski:2017ruc,Florkowski:2018fap,Yang:2018lew,Hattori:2019lfp,Fukushima:2020ucl,Shi:2020htn,Li:2020eon,Hongo:2021ona} in order to explore non-equilibrium effects. These studies have almost exclusively focused on the role of quarks, while the role of gluons in quantum chromodynamics (QCD) has been largely ignored. Such an investigation is the focus of this paper, in which we explore how the spin transport of fermions in the QGP is affected by the presence of gluons. Our formalism accounts for the possibility that dynamically generated color fields, originating from Weibel-type instabilities in an expanding QGP \cite{Mrowczynski:1988dz,Mrowczynski:1993qm,Romatschke:2003ms}, could have a dominant effect on transport properties of the QGP compared with collisional effects at weak coupling \cite{Asakawa:2006tc,Asakawa:2006jn}. 

Our investigation is closely linked to the conjectured phenomenon of local parity violation in QCD, where a locally fluctuating axial charge density could be produced by sphaleron transition in the QGP \cite{McLerran:1990de,Arnold:1996dy,Moore:2010jd} or by overpopulated gluons in the initial stage of the collision \cite{Mace:2016svc,Tanji:2016dka}. Such an axial charge density may also induce turbulent color fields via the chiral plasma instability \cite{Joyce:1997uy,Akamatsu:2013pjd,Mace:2019cqo}. The chiral magnetic effect \cite{Vilenkin:1980fu, Kharzeev:2007jp,Fukushima:2008xe}, where an electric current is induced by magnetic fields with local chirality imbalance, is a much studied transport phenomenon that, if observed, would reveal the presence of local parity violation. However, the observation of this effect is challenged by large backgrounds \cite{Koch:2016pzl,STAR:2021mii}. There have also been recent proposals to probe local parity violation characterized by an axial chemical potential with spin polarization in relativistic heavy ion collisions \cite{Becattini:2020xbh,Gao:2021rom}. Nevertheless, the axial charge evolves dynamically, and a dynamical study beyond an equilibrium assumption is needed. 

In this paper, we derive the transport theory giving rise to dynamical spin polarization of quarks and the non-vanishing axial-charge current induced by parity-odd color-field correlators in weakly coupled QGP. To track the spin transport of quarks in phase space, we extend the collisionless axial kinetic theory (AKT) to a QKT, in order to delineate the intertwined dynamics between charge and spin evolution of relativistic fermions in quantum electrodynamics (QED) \cite{Hattori:2019ahi,Yang:2020hri} to one for massive quarks in QCD. Following the approach in \cite{Asakawa:2006tc,Asakawa:2006jn} to extract the color-singlet component pertinent to physical observables, we derive the kinetic equation for spin containing not only a diffusion term but also a source term from odd-parity color field correlators which gives rise to dynamical spin polarization manifested in equilibrium. Such terms due to color-fields correlators are at $\mathcal{O}(g^2)$, where $g$ denotes the QCD coupling, which dominate over the expected collision term at $\mathcal{O}(g^4\ln g)$ at weak coupling \cite{Li:2019qkf,Yang:2020hri}. Furthermore, the explicit quantum corrections in color-singlet Wigner functions give rise to similar contributions upon spin polarization. The obtained quark polarization, accompanied with an axial-charge current, induced by parity-odd correlation of color fields might be regarded as an indirect signal for local parity violation in QGP. We finally discuss its phenomenological application to the spin alignment of vector mesons.  

\sect{Wigner functions and axial kinetic theory}
The AKT involves a scalar kinetic equation (SKE) and an axial-vector kinetic equation (AKE) obtained from the Wigner-function approach by expanding in $\hbar$. We will only highlight a few critical steps for the generalization to QCD. The Wigner transformation applied to quantum expectation values of correlation functions reads 
\begin{eqnarray}\label{eq:WT}
S^{<}(p,X)=\int d^4Ye^{ip\cdot Y/\hbar}\langle\bar{\psi}(y)U(y,x)\psi(x)\rangle,
\end{eqnarray}
where $X=(x+y)/2$, $Y=x-y$. Here $U(y,x)$ denotes the gauge link and $p_{\mu}$ represents the kinetic momentum, which ensure the gauge invariance of $S^{<}(p,X)$.
The Wigner function $S^{<}(p,X)$ is a matrix in color space.

The dynamical evolution of $S^{<}(p,X)$ is governed by the Kadanoff-Baym equation \cite{Elze:1986qd}, 
\begin{eqnarray}
	\bigg(\gamma^{\mu}\hat{\Pi}_{\mu}-m+\frac{i\hbar}{2}\gamma^{\mu}\hat{\nabla}_{\mu}\bigg)S^{<}=0,
\end{eqnarray}
where $\gamma^{\mu}$ are Dirac matrices. Up to order $\mathcal{O}(\hbar)$ the transport terms are given by 
\begin{eqnarray}\label{eq:def_Pi}
\hat{\Pi}_{\mu}S^{<} &\approx& p_{\mu}S^{<}+\frac{i\hbar}{8}[F_{\nu\mu},\partial_{p}^{\nu}S^{<}]_{},
\\
\hat{\nabla}_{\mu}S^{<} &\approx& D_{\mu}S^{<}+\frac{1}{2}\{F_{\nu\mu},\partial_{p}^{\nu}S^{<}\}_{}
\nonumber \\\label{eq:def_Delta}
&& -\frac{i\hbar}{24}[(\partial_{p}^{\rho} D_{\rho}F_{\nu\mu}),\partial_p^{\nu}S^{<}]_{},
\end{eqnarray}
where $D_{\mu}O=\partial_{\mu}O+i[A_{\mu},O]$ and the gauge coupling $g$ has been absorbed in the gauge field. Here square and curly brackets denote commutators and anti-commutators in color space, respectively. We will later decompose an arbitrary color object into $O=O^sI+O^at^a$, where $t^a$ are the SU($N_c$) generators and $I$ is the identity matrix. The collision term appears at order $\mathcal{O}(g^4\ln g)$ and is neglected. 

Following the standard approach \cite{Vasak:1987um}, we further decompose $S^{<}$ into its Clifford algebra components, where the only relevant components here are the vector and axial-vector, 
\begin{eqnarray}
\mathcal{V}^{\mu}(p,X) &=& \frac{1}{4}{\rm tr}\left(\gamma^{\mu}S^{<}(p,X)\right),
\\
\mathcal{A}^{\mu}(p,X) &=& \frac{1}{4}{\rm tr}\left(\gamma^{\mu}\gamma^{5}S^{<}(p,X)\right),
\end{eqnarray}
from which all other components can be derived. We further adopt the power counting, $\mathcal{V}^{\mu}\sim \mathcal{O}(\hbar^0)$ and $\mathcal{A}^{\mu}\sim \mathcal{O}(\hbar)$ due to the quantum nature of spin and retain only the leading-order contribution in the $\hbar$ expansion \cite{Yang:2020hri}. 

Perturbatively solving the Kadanoff-Baym equation at the considered order of $\hbar$ \cite{Hattori:2019ahi}, the vector Wigner function with the on-shell condition reads
\begin{equation}
\mathcal{V}^{\mu}=2\pi \delta(p^2-m^2)p^{\mu}f_V(p,X)
\end{equation}
and the axial-vector Wigner function has the form \cite{Hattori:2019ahi,Yang:2020hri}
\begin{eqnarray}\label{axial_sol}\nonumber
	\mathcal{A}^{\mu}&=&2\pi \Big[\delta(p^2-m^2)\Big(\tilde{a}^{\mu}+\hbar S^{\mu\nu}_{(n)}\tilde{\Delta}_{\nu}f_V\Big)
	\\
	&&+\frac{\hbar}{2} p_{\nu}\delta'(p^2-m^2)\{\tilde{F}^{\mu\nu},f_V\}_{}\Big],
\end{eqnarray} 
where $\tilde{\Delta}_{\mu}O=D_{\mu}O+\{F_{\nu\mu}\partial_{p}^{\nu},O\}/2$ and $f_V(p,X)$ denotes the vector-charge distribution function. The precise definition of $\tilde{a}^{\mu}(p,X)$ is given in Ref.~\cite{Hattori:2019ahi}. It is proportional to the spin four-vector delineating the orientation and distribution of spin polarization for massive fermions in phase space and satisfying the on-shell constraint $p\cdot \tilde{a}(p,X)=0$. In the massless limit, it reduces to $\tilde{a}^{\mu}(p,X)=p^{\mu}f_{A}(p,X)$ with $f_{A}(p,X)$ being the axial-charge distribution function. Here we also incorporate the magnetization current term proportional to $S^{\mu\nu}_{(n)}=\epsilon^{\mu\nu\alpha\beta}p_{\alpha}n_{\beta}/(2(p\cdot n+m))$, which depends on a frame vector $n^{\mu}$ based on the choice of spin bases \footnote{Note that the magnetization-current term is actually obtained by generalization from the case without background fields as in the derivation for QED.}, which does not affect physical results. The anti-commutators in color space emerge in (\ref{axial_sol}) as the primary difference compared to the case in QED. 

For convenience, we work from now on in the rest frame of massive quarks by choosing $n^{\mu}=n^{\mu}_{r}(p)\equiv p^{\mu}/m$ \cite{Weickgenannt:2019dks}, which is possible if $m$ is much larger than any gradients. We particularly focus on massive quarks since most of present experimental measurements are pertinent to spin polarization of strange quarks, but note that analogous results can also be derived for massless quarks \cite{Yang:2021fea}. In the rest frame, $S^{\mu\nu}_{m(n_r)}=0$ and hence the magnetization current term in (\ref{axial_sol}) vanishes. The rest part of the Kadanoff-Baym equation yields the SKE and AKE with full expressions shown in \cite{Yang:2021fea}, while their explicit forms are not physically illuminating. Instead, we will present the results after color decomposition and physical approximations. Moreover, we will focus on the off-shell kinetic equations and impose the on-shell conditions only when evaluating physical observables through the Wigner functions.   

We now decompose $\mathcal{V}^{\mu}$ and $\mathcal{A}^{\mu}$ into color singlet and octet components. In a weakly coupled QGP, the lowest-order contributions to singlet and octet quantitites are of order $\mathcal{O}(g^{0})$ and $\mathcal{O}(g)$, respectively. The derived AKE can thus be decomposed into off-shell transport equations in four-momentum space of the form
\begin{eqnarray}\label{AKE_SO}\nonumber
	\mathcal{K}_{\rm s}[\tilde{a}^{\mu}]+\hbar\mathcal{Q}^{\mu}_{\rm s}[f_V] &=& 0,
	\\
	\mathcal{K}_{\rm o}^a[\tilde{a}^{\mu}]+\hbar \mathcal{Q}^{a\mu}_{\rm o}[f_V] &=& 0,
\end{eqnarray}
where 
\begin{eqnarray}\nonumber
	\mathcal{K}_{\rm s}[O]&\equiv&p^{\mu}\Big(\partial_{\mu}O^{ s}+\bar{C}_2F^a_{\nu\mu}\partial_{p}^{\nu}O^{a}\Big),
	\\
	\mathcal{Q}^{\mu}_{\rm s}[O]&\equiv&
	-\frac{\bar{C}_2}{2}\epsilon^{\mu\nu\rho\sigma}p_{\rho}
	(\partial_{\sigma}F^a_{\beta\nu})\partial_{p}^{\beta}O^a,
\end{eqnarray}
with the force terms up to $\mathcal{O}(g^2)$ and $\bar{C}_2=1/(2N_c)$, and
\begin{eqnarray}\nonumber
	\mathcal{K}_{\rm o}^a[O]&\equiv&p^{\mu}\Big(\partial_{\mu}O^a-f^{bca}A^b_{\mu}O^c+F^a_{\nu\mu}\partial_{p}^{\nu}O^{\rm s}\Big),
	\\
	\mathcal{Q}^{a\mu}_{\rm o}[O]&\equiv &-\frac{1}{2}\epsilon^{\mu\nu\rho\sigma}p_{\rho}
	(\partial_{\sigma}F^a_{\beta\nu})\partial_{p}^{\beta}O^{s},
\end{eqnarray}
up to $\mathcal{O}(g)$ \footnote{We keep the $f^{bca}$ term responsible for the gauge link implicitly embedded in the color-field correlators introduced later}. The first and second equations in (\ref{AKE_SO}) correspond to the color-single and color-octet AKEs, respectively. Similarly, the color-singlet and color-octet SKEs are given by 
\begin{eqnarray}\label{SKE_SO}
\mathcal{K}_{\rm s}[f_V] = 0,
\quad 
\mathcal{K}^{a}_{\rm o}[f_V] = 0.
\end{eqnarray}
Note the absence of explicit spin terms in the SKEs at leading order in $\hbar$.  Since the coupled equations in the AKE and in the SKE are linear in $\tilde{a}^{\mu}$ and $f_V$, one can solve for  $f^a_{V}$ and $\tilde{a}^{a\mu}$ in terms of $f^s_{V}$ and $\tilde{a}^{s\mu}$.

\sect{Spin polarization and axial-charge currents} 
Here we are primarily interested in the momentum spectrum of spin polarization governed by $\mathcal{A}^{s\mu}(p,X)$. Using (\ref{axial_sol}) with $n^{\mu}=n^{\mu}_{r}(p)$ and expressing $f^{a}_{V}$ in terms of $f^{s}_V$ from (\ref{SKE_SO}) following Ref.~\cite{Asakawa:2006jn}, the axial charge current for quarks is given by
\begin{eqnarray}\nonumber\label{J5_def}
	J^{\mu}_5&=&4\int\frac{d^4p}{(2\pi)^4}{\rm Tr}_c\,\mathcal{A}^{\mu}(p,X)
	\\
	&=&4N_c\int\frac{d^4p}{(2\pi)^3}\delta(p^2-m^2)\big(\tilde{a}_{\rm s}^{\mu}+\hbar\bar{C}_2\mathcal{A}^{\mu}_{Q}\big)
	,
\end{eqnarray}
where ${\rm Tr}_c$ denotes the trace over colors. 
The polarization source term
\begin{equation}\label{AQmu_origin}
	\mathcal{A}^{\mu}_{Q}=\frac{\partial_{p\kappa}}{2}\int^{p}_{k,X'}p^{\beta}\langle \tilde{F}^{a\mu\kappa}(X)F^a_{\alpha\beta}(X')\rangle\partial_{p}^{\alpha}f^{\rm s}_V(p,X')
\end{equation}
is obtained from (\ref{axial_sol}) by integration by parts and dropping a surface term. Here we introduced the abbreviation
\begin{equation}
	\int^{p}_{k,X'}\equiv \int\frac{d^4k d^4X'}{(2\pi)^4}e^{ik\cdot(X'-X)}\left(\pi\delta(p\cdot k)+i\frac{{\cal P}}{p\cdot k}\right)
\end{equation}
with ${\cal P}$ denoting the principal value. We also introduced the correlator of color fields, which originates from the ensemble average of the gauge field and color-octet distribution function $\langle \tilde{F}^{a\mu\kappa}(X)f^a_V(p,X)\rangle$, assuming that the gauge field fluctuates much faster than the dynamical evolution of quarks. It would be of interest to test this assumption in numerical simulations of the full set of kinetic equations coupled to the Yang-Mills equation.

From (\ref{J5_def}) one can calculate the spin-polarization spectrum of massive quarks over the freeze-out hypersurface $\Sigma_{\mu}$ via
the modified Cooper-Frye formula \cite{Becattini2013a,Fang:2016vpj}, 
\begin{equation}\label{Spin_CooperFrye}
	\mathcal{P}^{\mu}({\bf p})=\frac{\int d\Sigma\cdot p\mathcal{J}_{5}^{\mu}(p,X)}{2m\int d\Sigma\cdot\mathcal{N}(p,X)}.
\end{equation}
Here (see Ref.~\cite{Yang:2021fea} for the detailed derivation)
\begin{eqnarray}
\mathcal{N}^{\mu}(p,X)&=& 4N_cp^{\mu}f^s_{V},
\\\label{eq:J5mu_density}
\mathcal{J}_5^{\mu}(p,X)&=&4N_c\big(\tilde{a}^{ s\mu}+\hbar\bar{C}_2(\mathcal{A}^{\mu}_{Q}+\mathcal{A}^{\mu}_{S})\big),
\end{eqnarray}
correspond to the on-shell number and axial-charge density currents in phase space, respectively, with the on-shell condition $p^0=\epsilon_{\bm p}\equiv \sqrt{|\bm p|^2+m^2}$. Particularly,
\begin{eqnarray}\nonumber
\mathcal{A}^{\mu}_{S}&=&-\frac{p_0}{2}\left(\partial_{pj}-\frac{p_{j}}{p_0}\partial_{p0}\right)\bigg[\int^{p}_{k,X'}\frac{p^{\beta}}{p_0}
\\
&&\times \langle \tilde{F}^{a\mu j}(X)F^a_{\alpha\beta}(X')\rangle\partial_{p}^{\alpha}f^{\rm s}_V(p,X')\bigg]
\end{eqnarray}
with $j=1,2,3$ corresponds to the contribution from the omitted surface term when computing the current, which has to be retrieved when evaluating the spectrum. The color-singlet distributions $\tilde{a}^{s\mu}$ and $f^s_{V}$ are found by solving the color-singlet kinetic equations (\ref{AKE_SO}) and (\ref{SKE_SO}). As already noted, the linearity of the equations allows us to express the color-octet distributions $\tilde{a}^{a\mu}$ and $f^a_{V}$ in terms of the singlet distributions $\tilde{a}^{s\mu}$ and $f^s_{V}$ via the color-octet kinetic equations and obtain a closed set of kinetic equations for the color-singlet distributions:
\begin{eqnarray}\label{AKE_singlet_simplify}
0&=&p\cdot\partial\tilde{a}^{s\mu}(p,X)-\partial_{p}^{\kappa}\mathscr{D}_{\kappa}[\tilde{a}^{s\mu}]
+\hbar\partial_{p}^{\kappa}\big(\mathscr{A}^{\mu}_{\kappa}[f^{\rm s}_{V}]\big),
\\\label{SKE_singlet_simplify}
0&=&p\cdot\partial f^{s}_V(p,X)-\partial_{p}^{\kappa}\mathscr{D}_{\kappa}[f_V^{s}],
\end{eqnarray}
where 
\begin{eqnarray}\nonumber
		\mathscr{D}_{\kappa}[O]&=&\bar{C}_2\int^{p}_{k,X'}p^{\lambda}p^{\rho}\langle F^a_{\kappa\lambda}(X)
		F^a_{\alpha\rho}(X')\rangle 
		\\
		&&\times\partial_{p}^{\alpha}O(p,X')
\end{eqnarray}
and
\begin{eqnarray}\nonumber
	\mathscr{A}^{\mu}_{\kappa}[O]&=&\frac{\bar{C}_2}{2}
	\epsilon^{\mu\nu\rho\sigma}\int^{p}_{k,X'}p^{\lambda}p_{\rho}\Big(\partial_{X'\sigma}\langle F^a_{\kappa\lambda}(X)F^a_{\alpha\nu}(X')\rangle
	\\
	&&+\partial_{X\sigma}\langle F^a_{\kappa\nu}(X)F^a_{\alpha\lambda}(X')\rangle\Big)\partial^{\alpha}_{p}O(p,X'),
\end{eqnarray}
where we introduce $\partial_{V\mu}\equiv \partial/\partial V^{\mu}$ and $\partial_{V}^{\mu}\equiv \partial/\partial V_{\mu}$ for an arbitrary $V^{\mu}$.
The equations (\ref{AQmu_origin}), (\ref{AKE_singlet_simplify}) rely on the assumption that the evolution of the gauge field can be approximately decoupled from the dynamical evolution of the quarks with the possible exception of screening effects that modify the soft gauge field sector. In a full dynamical scenario far off equilibrium, the gauge field correlators will be general functions of $X$ and $X'$. Since local equilibrium of gluons is thought to be rapidly attained in heavy ion collisions, at least within the same time scale for the thermalization of quarks \cite{Du:2020dvp}, we will proceed further by making specific assumptions about the form of these correlators.

\begin{figure}
	\begin{center}
		\includegraphics[width=0.75\hsize]{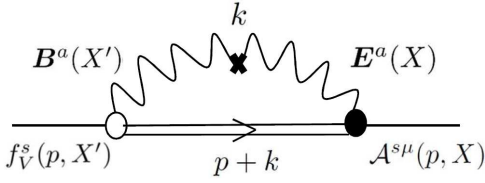}
	\end{center}
	\caption{An example for spin polarization triggered by the correlator of electric and magnetic gauge fields. The double line denotes an intermediate quark propagator connecting the unfilled and filled blobs, which correspond to the vertex of Lorentz force and of quantum correction mixing the charge and spin transport, respectively. The wiggly line with a cross at the center represents the correlator of color fields originating from the medium.
	}
	\label{fig:dynam_pol}
\end{figure}

Before proceeding further, we want to draw the reader's attention to the fact that, unlike the color-singlet SKE in (\ref{SKE_singlet_simplify}) which contains only a diffusion term $\mathscr{D}_{\kappa}[f_V^{s}]$, the equation (\ref{AKE_singlet_simplify}) for the color-singlet AKE contains a source term of the form $\hbar\mathscr{A}^{\mu}_{\kappa}[f^{\rm s}_{V}]$. This source term can generate dynamical spin polarization in addition to the explicit source term  $\mathcal{A}^{\mu}_{Q}+\mathcal{A}^{\mu}_{S}$ presented in (\ref{eq:J5mu_density}). Such induced spin polarization is possible because of the coupled dynamics of the color-singlet and octet components and between the charge and spin transport as illustrated in Fig.~\ref{fig:dynam_pol}.

\sect{Near-equilibrium scenario} 
The true gauge field correlators in the QGP can only be obtained from real-time simulations for prescribed initial conditions. To make further progress, we here assume an approximate form based on physical arguments. Assuming local space-time translational invariance of the QGP, we expect $\langle F^a_{\kappa\lambda}(X)F^a_{\alpha\rho}(X')\rangle  $ to depend only on the space-time separation $X-X'$. We further assume that all tensor components of the correlator depend on $X-X'$ in the same way:
\begin{equation}
\langle F^a_{\kappa\lambda}(X)F^a_{\alpha\rho}(X')\rangle = \langle F^a_{\kappa\lambda}F^a_{\alpha\rho}\rangle\Phi(X-X') ,
\end{equation} 
subject to the constraints imposed by the Bianchi identity. Since thermal gauge field fluctuations are Landau damped in the space-like domain, we further assume that the field correlator can be approximately represented by a time-like Gaussian as $\Phi(X-X')=e^{-(t-t')^2/\tau_c^2}$, where $\tau_c$ denotes the correlation time. This is in line with our focus on massive quarks that propagate along time-like trajectories, although this argument holds only marginally for strange quarks. In general, one may also account for spatial correlations without introducing principally new properties of the kinetic equations.

To make further progress, we introduce the electric and magnetic gauge fields in the QGP rest frame, $E^{a\mu}=F^{a\mu\nu}u_{\nu}$ and $B^{a\mu}=\epsilon^{\mu\nu\alpha\beta}u_{\nu}F^a_{\alpha\beta}/2$, where $u^{\mu}$ denotes the four-velocity of the thermal medium. We further work in the fluid-rest frame such that $u^{\mu}\approx (1,{\bm u})$ with $|\bm u|\ll 1$ and $p^{\mu}_{\perp}= p^{\mu}-u\cdot pu^{\mu}$. In perturbative QGP, static chromo-electric fields are screened but static chromo-magnetic fields remain unscreened \cite{Weldon:1982aq}. It is also well known that soft chromo-magnetic excitations in an expanding QGP are dynamically unstable against spontaneous amplification 
\cite{Mrowczynski:1993qm}. It is thus reasonable to assume that non-equilibrium correlators involving chromo-magnetic fields dominate over those involving chromo-electric fields with the hierarchy $|\langle B^{a}_{\mu}B^{a}_{\nu}\rangle| \gg |\langle E^{a}_{\mu}B^{a}_{\nu}\rangle| \gg |\langle E^{a}_{\mu}E^{a}_{\nu}\rangle|$. In thermal equilibrium at temperature $T$ and baryon chemical potential $\mu$, the color-singlet SKE is then satisfied by the Fermi distribution $f^{\rm s}_{V}(p)=f_{\rm eq}(p\cdot u)\equiv (e^{(p\cdot u-\mu)/T}+1)^{-1}$ , while the local fluctuations mediate anomalous dissipative transport processes \cite{Asakawa:2006jn}. 

In the absence of initial spin polarization, the diffusive term in the AKE (\ref{AKE_singlet_simplify}) is relatively negligible compared with the source term given by
\begin{eqnarray}\label{sorce_simpl}\nonumber
	\partial_{p}^{\kappa}\big(\mathscr{A}^{\mu}_{\kappa}[f^s_{V}]\big)_{\rm eq}
	&=&\frac{\bar{C}_2}{2p_0}
	\big(\partial_{p0}f_{\rm eq}(p_0)\big)\big(\langle B^{a\mu}E^{a\nu}\rangle p_{\nu}
	\\
	&&-\langle B^a\cdot E^a\rangle p^{\mu}_{\perp}\big),
\end{eqnarray}
which is actually contributed by $\Phi(0)$ and hence independent of $\tau_c$ as can be derived from integration by parts. 
The solution of (\ref{AKE_singlet_simplify}) for the initial condition $\tilde{a}^{\mu}(t_0,p)=0$ then is:
\begin{eqnarray}\label{amu_from_source}\nonumber
	\tilde{a}^{s\mu}(t,p)&=&-\frac{\hbar\bar{C}_2(t-t_0)}{2p_0^2}
	\big(\partial_{p0}f_{\rm eq}(p_0)\big)\big(\langle B^{a\mu}E^{a\nu}\rangle p_{\nu}
	\\
	&&-\langle B^a\cdot E^a\rangle p^{\mu}_{\perp}\big).
\end{eqnarray}
Additionally, we obtain for the explicit source term (\ref{eq:J5mu_density}): 
\begin{eqnarray}\nonumber\label{AQmu}
(\mathcal{A}^{\mu}_{Q})_{\rm eq}+(\mathcal{A}^{\mu}_{S})_{\rm eq}&\approx&\frac{\sqrt{\pi}\tau_c}{4\epsilon_{\bm p}^3}\Big[(p^{\alpha}p^{\beta}\langle E^a_{\alpha}B^a_{\beta}\rangle u^{\mu}
-\langle B^{a\mu}E^{a\nu}\rangle
\\
&&\times \epsilon_{\bm p}(\epsilon_{\bm p}^2\partial_{p\nu}-p_{\nu})\Big]
\partial_{\epsilon_{\bm p}}f_{\rm eq}(\epsilon_{\bm p}).
\end{eqnarray}
The two terms, which contribute to the axial-charge density current (\ref{eq:J5mu_density}) via $\mathcal{A}^{s\mu}$, differ in notable aspects. The source term in (\ref{amu_from_source}) is secular, while that in (\ref{AQmu}) is limited by the correlation time of gauge field fluctuations. Conversely, the contribution from $\tilde{a}^{\mu}_{s}$ is relatively suppressed in the large-mass limit due to an extra $1/p_0$ factor. 

Given $\mathcal{A}^{s\mu}$, we may further evaluate the axial-charge current, 
\begin{eqnarray}\nonumber\label{J5_equilibrium}
	J^{\mu}_5&=&4N_c\int\frac{d^4p}{(2\pi)^4}{\rm sign}(p_0)\mathcal{A}^{s\mu}(p,X)
\\
&=&-\frac{\hbar u^{\mu}}{8\pi^2}\sqrt{\pi}\tau_c
\langle E^a\cdot B^a\rangle \mathcal{I},
\end{eqnarray}
where $\mathcal{I}=\int^{\infty}_0\frac{d|\bm p||\bm p|^2}{T\epsilon_{\bm p}^2}\big(\mathcal{F}(\epsilon_{\bm p})+\mathcal{F}(-\epsilon_{\bm p})\big)$ with $\mathcal{F}(\epsilon_{\bm p})=f_{\rm eq}(\epsilon_{\bm p})\big(1-f_{\rm eq}(\epsilon_{\bm p})\big)$,
which includes the contribution from anti-quarks. 

This axial charge in the medium rest frame only derives from $\mathcal{A}^{\mu}_Q$ and is constant, thus yields $\partial\cdot J_{5}=0$. Diagrammatically, the vanishing axial Ward identity here corresponds to a triangle diagram with two gluon legs connected and zero momentum flow from the axial vertex. The constant nonvanishing axial charge is analogous to the one $\sim \tau_{R}E\cdot B$ produced in Weyl semimetals from parallel electric and magnetic fields by balancing the chiral anomaly and the inter-cone interaction in a steady state such that $\partial\cdot J_5=0$, where $\tau_R$ represents the inter-cone relaxation time \cite{Son:2012bg,Li:2014bha}. 

As mentioned previously, one can derive similar results for massless quarks \cite{Yang:2021fea}, applicable to up and down quarks in QCD, by employing the chiral kinetic theory (CKT) \cite{Stephanov:2012ki,Son:2012wh,Chen:2014cla} in the Wigner function formalism \cite{Chen:2012ca,Hidaka:2016yjf,Hidaka:2017auj,Huang:2018wdl,Luo:2021uog}. Although the kinetic equations and the dynamics of spin polarization for massless quarks are more complicated than in the massive case due to the involvement of side-jump terms originating from the spin-orbit interaction \cite{Chen:2014cla, Chen:2015gta,Hidaka:2016yjf,Yang:2018lew}, the general structure remains unchanged. Furthermore, near thermal equilibrium, it turns out that only $\mathcal{A}^{\mu}_{Q}$ taking the same form as (\ref{AQmu}) contributes to $\mathcal{J}^{\mu}_5$. One also recovers the same $J^{\mu}_5$ as in (\ref{J5_equilibrium}) with $\mathcal{I}=1$ for $m=0$. This means that the source term for the axial current density is approximately given by (\ref{AQmu}) both in the limit $m=0$ and $m\gg |{\bm p}|$ (applicable for strange quarks with low momenta and constituent quark mass).

\textit{Spin alignments of vector mesons}.---
Based on our approximations and ignoring other contributions such as vorticity and gradient terms near equilibrium, the polarization $\mathcal{P}^{\mu}(\bm p)$ generated by $\mathcal{A}^{s\mu}(p)$ is the same for quarks and anti-quarks at zero chemical potential since $\mathcal{A}^{s\mu}(p)$ is charge-conjugation even. However, since the  gauge fields correlators fluctuate event by event in relativistic heavy ion collisions, the sign of $\mathcal{P}^{\mu}({\bf p})$ for quarks also fluctuates. Therefore, the contribution from the interaction with gauge fields gives a vanishing contribution to the global polarization of $\Lambda$ hyperons predominantly carried mainly by a single strange quark. On the other hand, it can contribute to the spin alignment of vector mesons originating from polarization of a quark and an anti-quark.  

In Ref.~\cite{Liang:2004xn}, it was found that the longitudinal (00) element of the spin density matrix of the vector meson is related to the polarization of a quark and an anti-quark through $\rho_{00}=(1-\mathcal{P}^{i}_{q}\mathcal{P}^{i}_{\bar{q}})/(3+\mathcal{P}^{i}_{q}\mathcal{P}^{i}_{\bar{q}})$.  Thus, $\rho_{00}\neq 1/3$ measured by the ALICE experiment for $K^{*0}$ and $\phi$ mesons \cite{ALICE:2019aid,Singha:2020qns} implies the spin polarization of QGP, yet the underlying mechanisms is unknown so far \cite{Sheng:2019kmk,Sheng:2020ghv,Xia:2020tyd}. We may employ (\ref{Spin_CooperFrye}) to evaluate $\mathcal{P}^{i}_{q/\bar{q}}$. Because $\mathcal{P}^{i}_{q}=\mathcal{P}^{i}_{\bar{q}}$ and $|\mathcal{P}^i_{q/\bar{q}}|$ is larger for light quarks, we expect $\rho_{00}(K^{*0})< \rho_{00}(\phi)<1/3$. This qualitative behavior is consistent with the ALICE measurements \cite{ALICE:2019aid,Singha:2020qns}. However, at RHIC it was found that $\rho_{00}(K^{*0})<1/3$ and $\rho_{00}(\phi)>1/3$ \cite{Singha:2020qns}. If confirmed, this result for the $\phi$ meson alignment may indicate that the contributions from other mechanisms are more prominent in lower energy collisions.   

\textit{Summary}.---
In conclusion, we have shown that parity-odd correlators of color fields can locally engender dynamical spin polarization and axial currents in a weakly coupled QGP. We further argued that this mechanism can qualitatively explain the spin alignment of vector mesons measured in high-energy nuclear collisions. Our findings reveal the possible significant influence of gluons on spin transport of quarks in heavy ion collisions and its implication on local parity violation in the QGP.
 
\acknowledgments
\textit{Acknowledgments}.
B. M. was supported by the U. S. Department of Energy under Grant No. DE-FG02-05ER41367.
D.-L. Y. was supported by Ministry of Science and Technology, Taiwan under Grant No. MOST 110-2112-M-00l-070-MY3.

\bibliography{color_field_Letter_arXiv_corrected.bbl}

\end{document}